\documentstyle[multicol,aps,psfig]{revtex}
\renewcommand{\narrowtext}{\begin{multicols}{2}
\global\columnwidth20.5pc\noindent}
\renewcommand{\widetext}{\end{multicols}
\global\columnwidth42.5pc}
\begin{document}
\draft
\preprint{13 December 2002}
\title{Spin-Wave Description of Haldane-gap antiferromagnets}
\author{Shoji Yamamoto and Hiromitsu Hori}
\address{Division of Physics, Hokkaido University,
         Sapporo 060-0810, Japan}
%\date{Received \hspace{4cm}}
\date{Received 13 December 2002}
\maketitle
\begin{abstract}
Modifying the conventional antiferromagnetic spin-wave theory which
is plagued by the difficulty of the zero-field sublattice
magnetizations diverging in one dimension, we describe magnetic
properties of Haldane-gap antiferromagnets.
The modified spin waves, constituting a grand canonical bosonic
ensemble so as to recover the sublattice symmetry, not only depict
well the ground-state correlations but also give useful information on
the finite-temperature properties.
\end{abstract}
\pacs{PACS numbers: 75.10.Jm, 05.30.Jp, 75.40.Mg}
% 75.10.Jm: Quantized spin models
%05.30.Jp: Boson systems
% 75.40.Mg: Numerical simulation studies
% 76.50.$+$g: Ferromagnetic, antiferromagnetic, and ferrimagnetic
%             resonances; spin-wave resonance
%75.50.Xx: Molecular magnets
\narrowtext

   Haldane's conjecture \cite{H464,H1153} that one-dimensional
Heisenberg antiferromagnets should exhibit qualitatively different
low-energy structures based on whether the constituent spins are integral
or fractional sparked renewed interest in the field of quantum
magnetism.
An energy gap immediately above the ground state was indeed observed
in a quasi-one-dimensional Heisenberg antiferromagnet
Ni(C$_2$H$_8$N$_2$)$_2$NO$_2$(ClO$_4$) \cite{R945} and a rigorous
example of such a massive phase was also found out
\cite{A799,A477}.
The energy gaps in magnetic excitation spectra, that is, spin gaps,
are now one of the most attractive and important topics.
In the context of theoretical progress, we may be reminded of quantized
plateaux in the ground-state magnetization curves \cite{O1984}, a dramatic
crossover from one- to two-dimensional quantum antiferromagnets
\cite{D618}, and an antiferromagnetic excitation gap accompanied by
ferromagnetic background \cite{M68,Y14008,Y11033,T15189}.
From the experimental point of view, metal oxides such as
spin-Peierls compounds
Cu$_{1-x}M_x$GeO$_3$ ($M=\mbox{Zn},\mbox{Mg}$) \cite{H3651,H4059},
Haldane-gap antiferromagnets
$R_2$BaNiO$_5$ ($R=\mbox{rare\ earth}$) \cite{D409} and
ladder materials
Sr$_{n-1}$Cu$_{n+1}$O$_{2n}$ ($n=3,5,7,\cdots$) \cite{A3463}
have significantly contributed to systematic
investigations of the mechanism of gap formation.

   In studying Haldane-gap antiferromagnets, numerical tools such as
quantum Monte Carlo and density-matrix renormalization-group methods are
indeed useful, but analytic approaches still play an important role.
The nonlinear $\sigma$-model technique
\cite{H464,H1153,A397,A409,F14709,F8799} is powerful enough to investigate
low-energy structures with particular emphasis on the competition
between massive and massless phases.
It is the valence-bond-solid description \cite{A799,A477} of
integer-spin chains that enables us to readily understand the hidden
order \cite{N4709} inherent in the Haldane massive phase.
However, these methods are not so useful for exploring thermal properties
as they are in investigating the ground-state properties.
Although we may consider applying the spin-wave scheme to
Haldane-gap antiferromagnets in such circumstances, the difficulty of
the zero-field sublattice magnetizations diverging in one dimension
has been a problem in the conventional spin-wave theory \cite{A694,K568}
for years.
Therefore, it was a major breakthrough that spin waves, being constrained
to keep zero magnetization, succeeded in precisely describing the
low-temperature thermodynamics of the one-dimensional Heisenberg
ferromagnet \cite{T168}.
This new spin-wave scheme, which is now referred to as the modified
spin-wave theory, was further applied to quantum antiferromagnets and
ferrimagnets.
The modified spin-wave scheme is highly successful for extensive
ferrimagnets \cite{Y211,Y1,N1380,N214418,Y157603,O8067} and still applies
well for two-dimensional antiferromagnets
\cite{T2494,H4769,H5000,C7832,D13821}.
As for one-dimensional antiferromagnets, there exists a pioneering
argument \cite{R2589}, but it looks unsatisfactory for interpreting
experimental and numerical observations.

    In this article, we aim to do away with our vague but persistent
impression that the spin-wave scheme hardly works for one-dimensional
quantum antiferromagnets.
Spin-$2$ Haldane-gap antiferromagnets \cite{G1616,Y6831}, as well as
those of spin $1$, have recently been synthesized and more explorations
into novel quantum phenomena in one dimension are expected in the future.
In such circumstances, we make our first attempt to construct a
{\it modified spin-wave theory for Haldane-gap antiferromagnets}.

   We consider integer-spin antiferromagnetic Heisenberg chains
\begin{equation}
   {\cal H}
   =J\sum_{j=1}^L
    \mbox{\boldmath$S$}_j\cdot\mbox{\boldmath$S$}_{j+1}\,;\ \ 
    \mbox{\boldmath$S$}_j^2=S(S+1)\,.
   \label{E:H}
\end{equation}
We define bosonic operators for the spin deviation in each sublattice via
\begin{equation}
   \left.
   \begin{array}{lll}
      S_{2n-1}^+=\sqrt{2S-a_n^\dagger a_n}\ a_n\,,&
      S_{2n-1}^z=S-a_n^\dagger a_n\,,\\
      S_{2n}^+=b_n^\dagger\sqrt{2S-b_n^\dagger b_n}\ ,&
      S_{2n}^z=-S+b_n^\dagger b_n\,.
   \end{array}
   \right.
   \label{E:HP}
\end{equation}
The Fourier-transformed operators are introduced as
\begin{equation}
   \left.
   \begin{array}{l}
      a_k={\displaystyle\frac{1}{\sqrt{N}}}
          {\displaystyle\sum_{n=1}^N}
          {\rm e}^{ {\rm i}k(n-1/4)}a_n\,,\\
      b_k={\displaystyle\frac{1}{\sqrt{N}}}
          {\displaystyle\sum_{n=1}^N}
          {\rm e}^{-{\rm i}k(n+1/4)}b_n\,,
   \end{array}
   \right.
   \label{E:FT}
\end{equation}
where twice the lattice constant $2a$ is set equal to unity and therefore
$k=2\pi n/N\ (n=0,1,\cdots,N-1;\,N=L/2)$.
Through the Bogoliubov transformation
\begin{equation}
   \left.
   \begin{array}{lll}
      a_k^\dagger&=&
      \alpha_k^\dagger{\rm cosh}\theta_k-\beta_k{\rm sinh}\theta_k\,,\\
      b_k^\dagger&=&
      \beta_k^\dagger{\rm cosh}\theta_k-\alpha_k{\rm sinh}\theta_k\,,
   \end{array}
   \right.
   \label{E:BT}
\end{equation}
we obtain
\begin{equation}
   {\cal H}=E_{\rm N}+E_1+E_0+{\cal H}_1+{\cal H}_0+O(S^{-1})\,,
   \label{E:HSW}
\end{equation}
where
\begin{mathletters}
   \begin{eqnarray}
      E_{\rm N}&=&-2NJS^2\,,\\
      E_1&=&-4NJS({\mit\Gamma}-{\mit\Lambda})\,,\\
      E_0&=&-2NJ({\mit\Gamma}-{\mit\Lambda})^2\,,
   \end{eqnarray}
\end{mathletters}
\vspace*{-6mm}
\begin{mathletters}
   \begin{eqnarray}
      {\cal H}_1
       &=&J\sum_k
          \Bigl[
           \omega_1(k)\left(\alpha_k^\dagger\alpha_k
                           +\beta_k^\dagger\beta_k\right)
       \nonumber\\
       & &\qquad
          +\gamma_1(k)\left(\alpha_k\beta_k
                           +\alpha_k^\dagger\beta_k^\dagger\right)
          \Bigr]\,,\\
      {\cal H}_0
       &=&J\sum_k
          \Bigl[
           \omega_0(k)\left(\alpha_k^\dagger\alpha_k
                           +\beta_k^\dagger\beta_k\right)
       \nonumber\\
       & &\qquad
          +\gamma_0(k)\left(\alpha_k\beta_k
                           +\alpha_k^\dagger\beta_k^\dagger\right)
          \Bigr]\,,
      \end{eqnarray}
\end{mathletters}
with
\begin{mathletters}
   \begin{eqnarray}
   {\mit\Gamma}
    &=&\frac{1}{2N}\sum_k\cos\frac{k}{2}\,{\rm sinh}2\theta_k\,,\\
   {\mit\Lambda}
    &=&\frac{1}{2N}\sum_k({\rm cosh}2\theta_k-1)\,,
   \end{eqnarray}
\end{mathletters}
\vspace*{-6mm}
\begin{mathletters}
   \begin{eqnarray}
   \omega_1(k)
    &=&2S\left({\rm cosh}2\theta_k
              -\cos\frac{k}{2}\,{\rm sinh}2\theta_k\right)\,,\\
   \omega_0(k)
    &=&2({\mit\Gamma}-{\mit\Lambda})
       \left({\rm cosh}2\theta_k
            -\cos\frac{k}{2}\,{\rm sinh}2\theta_k\right)\,,
   \end{eqnarray}
\end{mathletters}
\vspace*{-6mm}
\begin{mathletters}
   \begin{eqnarray}
   \gamma_1(k)
    &=&2S\left(\cos\frac{k}{2}\,{\rm cosh}2\theta_k
              -{\rm sinh}2\theta_k\right)\,,\\
   \gamma_0(k)
    &=&2({\mit\Gamma}-{\mit\Lambda})
       \left(\cos\frac{k}{2}\,{\rm cosh}2\theta_k
            -{\rm sinh}2\theta_k\right)\,.
      \end{eqnarray}
\end{mathletters}
The naivest diagonalization of the Hamiltonian (\ref{E:HSW}), whether
up to $O(S^1)$ or up to $O(S^0)$, results in diverging sublattice
magnetizations even at zero temperature.
Although eq. (\ref{E:HP}) assumes that $a_n^\dagger a_n\leq 2S$ and
$b_n^\dagger b_n\leq 2S$, the conventional spin-wave theory cannot
reasonably control the boson numbers.
Then we consider introducing a grand canonical constraint to the
noncompact Hamiltonian (\ref{E:HSW}).

   Isotropic magnets should lie in the state of zero magnetization
$\sum_j S_j^z=0$ and the minimization of the free energy under such
a condition indeed yields an excellent description of the
low-temperature thermodynamics of ferromagnets \cite{T168}.
However, in the cases with antiferromagnetic exchange interactions,
the zero-magnetization constraint, claiming that
$\sum_n(a_n^\dagger a_n-b_n^\dagger b_n)=0$, still fails to overcome
the divergence of the numbers of the sublattice bosons.
Hence we minimize the free energy {\it constraining the sublattice
magnetizations to be zero}:
\begin{equation}
   \sum_n a_n^\dagger a_n=\sum_n b_n^\dagger b_n=SN\,.
   \label{E:const}
\end{equation}
Within the conventional spin-wave theory, spins on one sublattice point
predominantly up, while those on the other predominantly down.
{\it This condition (\ref{E:const}) restores the sublattice symmetry.}
In order to enforce the constraint (\ref{E:const}), we first
introduce a Lagrange multiplier and diagonalize
\begin{equation}
   \widetilde{\cal H}
   ={\cal H}+2J\lambda\sum_k(a_k^\dagger a_k+b_k^\dagger b_k)\,.
   \label{E:effHSW}
\end{equation}
Then the ground-state energy and the dispersion relation are obtained as
\begin{eqnarray}
   &&
   E_{\rm g}=E_{\rm N}+\widetilde{E}_1\,;\ \ 
   \widetilde{E}_1=E_1+4NJ\lambda\Lambda\,,
   \label{E:LEg}\\
   &&
   \omega(k)=\widetilde{\omega}_1(k)\,;\ \ 
   \widetilde{\omega}_1(k)
    =\omega_1(k)+2\lambda{\rm cosh}2\theta_k\,,
   \label{E:Ldsp}
\end{eqnarray}
within the linear modified spin-wave scheme and as
\begin{eqnarray}
   &&
   E_{\rm g}=E_{\rm N}+\widetilde{E}_1+E_0\,,
   \label{E:IEg}\\
   &&
   \omega(k)=\widetilde{\omega}_1(k)+\omega_0(k)\,,
   \label{E:Idsp}
\end{eqnarray}
within the up-to-$O(S^0)$ interacting modified spin-wave scheme.
For eqs. (\ref{E:LEg}) and (\ref{E:Ldsp}), $\theta_k$ is given by
$\gamma_1(k)-2\lambda{\rm sinh}2\theta_k
 \equiv\widetilde{\gamma}_1(k)=0$,
whereas for eqs. (\ref{E:IEg}) and (\ref{E:Idsp}), $\theta_k$ may be
determined in two ways.
One idea is the perturbational treatment of ${\cal H}_0$, which is
referred to as the perturbational interacting modified spin-wave scheme,
where $\theta_k$ is still given by $\widetilde{\gamma}_1(k)=0$ and the
$O(S^0)$ quantum correction is the $O(S^1)$-eigenstate average of
${\cal H}_0$.
The other is the full diagonalization of ${\cal H}_1+{\cal H}_0$,
which is referred to as the full-diagonalization interacting modified
spin-wave scheme, where $\theta_k$ is given by
$\widetilde{\gamma}_1(k)+\gamma_0(k)=0$.
Once $\theta_k$ is given, we calculate the free energy and obtain the
optimum thermal distribution functions as
\begin{equation}
   \langle \alpha_k^\dagger\alpha_k\rangle
  =\langle \beta_k^\dagger \beta_k\rangle
   \equiv {\bar n}_k
  =\frac{1}{{\rm e}^{J\omega_k/k_{\rm B}T}-1}\,,
\end{equation}
where $\lambda$ is self-consistently determined by the condition
\begin{equation}
   \sum_k\left(2{\bar n}_k+1\right){\rm cosh}2\theta_k=(2S+1)N\,.
\end{equation}

   First, let us evaluate the ground-state energy.
We compare the modified spin-wave calculations with the highly accurate 
quantum Monte Carlo estimates \cite{T047203} in Table \ref{T:Eg}.
The modified spin-wave findings are generally in good agreement with the
quantum Monte Carlo results.
The interacting modified spin waves describe the ground-state correlations
much better than the linear ones and their description becomes
increasingly refined with increasing $S$.
The interacting modified spin-wave findings miss the correct value by 0.5
percent for $S=1$ and by only 0.008 percent for $S=3$.

\begin{figure}
\centerline
{\mbox{\psfig{figure=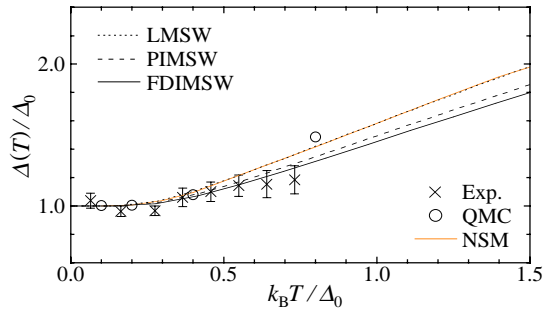,width=72mm,angle=0}}}
%\vspace*{1mm}
\caption{Temperature dependences of the spin-$1$ Haldane gap
         calculated by the linear (LMSW), perturbational interacting
         (PIMSW) and full-diagonalization interacting (FDIMSW)
         modified spin-wave schemes, a quantum Monte Carlo (QMC) method
         combined with the maximum-entropy technique [39] and the
         nonlinear $\sigma$-model (NSM) approach [40].
         Inelastic neutron-scattering measurements (Exp.) on
         Y$_2$BaNiO$_5$ [41] are also shown for reference.}
\label{F:gapT}
\end{figure}

   Secondly, we consider the Haldane gap
$\omega(ak=\pi)\equiv{\mit\Delta}(T)$.
Table \ref{T:gap} shows that the modified spin-wave scheme, in contrast
with the conventional spin-wave theory, succeeds in generating the gap but
considerably underestimates it.
Unavailability of the absolute energy level structure is an inevitable
consequence of our employing the {\it effective Hamiltonian}
(\ref{E:effHSW}).
Then we examine the present scheme by scaling ${\mit\Delta}(T)$ to its
zero-temperature value ${\mit\Delta}(T=0)\equiv{\mit\Delta}_0$.
Such an argument is quite usual with field-theoretical investigations,
which do not lead to an estimate of the normalization factor but derive
finite-temperature expressions involving only ratios such as
$k_{\rm B}T/{\mit\Delta}_0$.
In Fig. \ref{F:gapT}, we compare the modified spin-wave calculations with
numerical \cite{D10227}, field-theoretical \cite{J9265} and experimental
\cite{S3025} findings.
Now the superiority of the modified spin-wave scheme is clear at a glance.
Of all the theoretical tools, the full-diagonalization
interacting modified spin-wave approach is the most successful to
reproduce the observed upward behavior of ${\mit\Delta}(T)$ with
increasing temperature \cite{S3025,R3538,T4677}.
The nonlinear $\sigma$-model treatment \cite{J9265,A474} is justified well
in the low-temperature region $k_{\rm B}T\ll{\mit\Delta}_0$, while the
maximum-entropy technique \cite{D10227} works less with increasing
temperature.
Nickel compounds such as Y$_2$BaNiO$_5$ \cite{S3025} and
Ni(C$_2$H$_8$N$_2$)$_2$NO$_2$(ClO$_4$) \cite{R3538} are good candidates
for spin-$1$ Haldane-gap antiferromagnets, but magnetic anisotropy and
interchain interaction are not negligible there.
They split the lowest excitation gap of the isotropic chain
(\ref{E:H}) into several levels and smear the intrinsic behavior of
the ideal integer-spin chains.
Considering such practical factors, the present theory satisfactorily
interprets the observations.

\begin{figure}
\centerline
{\mbox{\psfig{figure=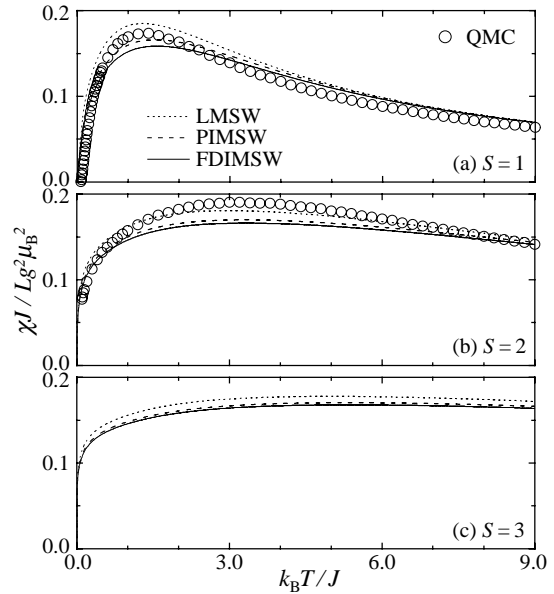,width=72mm,angle=0}}}
\vspace*{0mm}
\caption{The linear (LMSW), perturbational interacting (PIMSW) and
         full-diagonalization interacting (FDIMSW) modified spin-wave
         calculations of the zero-field magnetic susceptibility as a
         function of temperature compared with quantum Monte Carlo
         (QMC) estimates.}
\label{F:chi}
\end{figure}

   Lastly, we show the modified spin-wave calculations of the magnetic
susceptibility in Fig. \ref{F:chi}.
Considering the total breakdown of the conventional spin-wave theory in
one-dimensional thermodynamic calculations, the modified spin-wave
achievement is highly successful.
All the calculations are guaranteed to reproduce the paramagnetic
susceptibility
$\chi/Lg^2\mu_{\rm B}^2=S(S+1)/3k_{\rm B}T$ at high temperatures.
Since the interacting modified spin-wave scheme gives better estimates of
the gap than the linear modified spin-wave one (Table \ref{T:Eg}), it is
somewhat better at describing the low-temperature behavior.
With increasing $S$, the activation-type initial behavior is
suppressed and the antiferromagnetic peak is broadened.
The $S=1$ modified spin-wave calculations are in fine agreement with the
quantum Monte Carlo findings over a wide temperature range, while those
for $S=2$ look somewhat poorer at intermediate temperatures.
It may be closely related to the fact that the modified spin-wave
estimates of the $S=2$ gap are worse than those of the $S=1$ gap
(Table \ref{T:gap}).
However, Table \ref{T:gap} suggests that the validity of the modified
spin-wave scheme for excitations significantly improves with increasing
$S$, possibly in a staggered way.
Thermodynamic calculations, whether by quantum Monte Carlo or
density-matrix renormalization group, for systems with large degrees of
freedom are less feasible numerically, in particular, at low temperatures.
The present scheme has the advantage of saving time and computational
resources.

   We have demonstrated the applicability of the new spin-wave scheme to
Haldane-gap antiferromagnets.
{\it This is the first comprehensive attempt to describe one-dimensional
spin-gapped antiferromagnets in terms of spin waves}.
The modified spin waves are free from their thermal as well as quantum
divergence and can therefore microscopically interpret various magnetic
properties.
Besides the magnetic susceptibility, the spin correlation function and the
nuclear spin-lattice relaxation time can be revealed.
Unfortunately, we have less information on the bare energy spectrum,
because we get rid of the quantum divergence at the cost of the original
Hamiltonian.
However, with the zero-temperature spin gap, which is readily and
precisely available through numerical calculations within a canonical
ensemble \cite{T047203}, we can still argue the energy structure
quantitatively.
\vspace*{2mm}
\vspace*{-3mm}

   Haldane-gap antiferromagnets in a magnetic field provide further
interesting issues.
With the increase of an applied field, the gap is reduced and the ground
state is mixed increasingly with the first excited state.
Indeed the nuclear spin-lattice relaxation is accelerated in the vicinity
of the critical field, but its overall behavior as a function of an
applied field and temperature \cite{F11860} is far from understandable at
a glance.
We may expect the present new scheme to open the way for the total
understanding of low-dimensional spin-gapped antiferromagnets.

%   This work was supported by the Ministry of Education, Culture, Sports,
%Science and Technology of Japan through Grant-in-Aid No. 13740188 and by
%the Sumitomo Foundation.

\vspace{-2mm}
\begin{table}
\caption{The linear (LMSW), perturbational interacting (PIMSW) and
         full-diagonalization interacting (FDIMSW) modified spin-wave
         calculations of the ground-state energy per site compared
         with quantum Monte Carlo (QMC) estimates [38].}
\begin{tabular}{llll}
 & \quad\ $S=1$ & \quad\ $S=2$ & \quad$S=3$ \\
%\noalign{\vskip 1mm}
\tableline
\noalign{\vskip 1mm}
LMSW   & $-1.361879   $ & $-4.726749   $ & $-10.0901   $ \\
PIMSW  & $-1.394853   $ & $-4.759760   $ & $-10.1231   $ \\
FDIMSW & $-1.394617   $ & $-4.759759   $ & $-10.1231   $ \\
QMC    & $-1.401481(4)$ & $-4.761249(6)$ & $-10.1239(1)$ \\
\end{tabular}
\label{T:Eg}
\end{table}
\vspace*{-2mm}
\begin{table}
\caption{The linear (LMSW), perturbational interacting (PIMSW) and
         full-diagonalization interacting (FDIMSW) modified spin-wave
         calculations of the lowest excitation gap
         ${\mit\Delta}(T=0)\equiv{\mit\Delta}_0$ compared with
         quantum Monte Carlo (QMC) estimates [38].}
\begin{tabular}{llll}
 & \ \ $S=1$ & \ \ $S=2$ & \ \ $S=3$ \\
%\noalign{\vskip 1mm}
\tableline
\noalign{\vskip 1mm}
LMSW   & $0.07200   $ & $0.00626   $ & $0.00279   $ \\
PIMSW  & $0.07853   $ & $0.00655   $ & $0.00287   $ \\
FDIMSW & $0.08507   $ & $0.00683   $ & $0.00295   $ \\
QMC    & $0.41048(6)$ & $0.08917(4)$ & $0.01002(3)$ \\
\end{tabular}
\label{T:gap}
\end{table}

\widetext
\end{document}